# Cluster Language Model for Improved E-Commerce Retrieval and Ranking: Leveraging Query Similarity and Fine-Tuning for Personalized Results

**Duleep Rathgamage Don**
School of Data Science and Analytics
Kennesaw State University, USA
drathgam@students.kennesaw.edu

**Ying Xie**
Department of Information Technology
Kennesaw State University, USA
yxie2@kennesaw.edu

**Le Yu**
The Home Depot, USA
le_yu1@homedepot.com

**Simon Hughes**
The Home Depot, USA
simon_hughes@homedepot.com

**Yun Zhu**
The Home Depot, USA
yun_zhu@homedepot.com

## Abstract

This paper proposes a novel method to improve the accuracy of product search in e-commerce by utilizing a cluster language model. The method aims to address the limitations of the bi-encoder architecture while maintaining a minimal additional training burden. The approach involves labeling top products for each query, generating semantically similar query clusters using the K-Means clustering algorithm, and fine-tuning a global language model into cluster language models on individual clusters. The parameters of each cluster language model are fine-tuned to learn local manifolds in the feature space efficiently, capturing the nuances of various query types within each cluster. The inference is performed by assigning a new query to its respective cluster and utilizing the corresponding cluster language model for retrieval. The proposed method results in more accurate and personalized retrieval results, offering a superior alternative to the popular bi-encoder based retrieval models in semantic search.

## 1  Introduction

E-commerce platforms have experienced tremendous growth in recent years, with millions of users browsing and purchasing products online every day. One of the critical factors that contribute to a successful e-commerce platform is the ability to effectively retrieve and rank products based on user queries. A robust retrieval and ranking system should be able to understand the underlying semantics of queries and provide personalized results that meet the specific needs of individual users.

Traditional retrieval models, such as the Vector Space Model and Latent Semantic Analysis, have been effective in capturing keyword-based relevance between queries and items. However, they often struggle with understanding the nuances of natural language and user intent. Recent advancements in natural language processing and deep learning have led to the development of powerful pre-trained language models, such as BERT, GPT, and RoBERTa. These models have demonstrated impressive performance in various tasks, including e-commerce retrieval and ranking, by capturing the semantic relationships between queries and items.

Despite the successes of pre-trained language models, there is still room for improvement, particularly in understanding the diverse nature of user queries and providing tailored retrieval results. One promising direction is to incorporate query clustering into the retrieval and ranking process, leveraging the inherent structure of the query space to better adapt the model to different query types. Query clustering can help uncover underlying patterns in user search behavior and create more fine-grained representations of user intent, ultimately leading to improved retrieval performance

In this paper, we propose a cluster-based language model for e-commerce retrieval and ranking that builds upon the strengths of pre-trained language models and query clustering. Our method (Figure 2) first fine-tunes a pre-trained language model, on query-product pairs using a bi-encoder approach, forming a baseline model (Figure 1). We then cluster the training queries into k clusters and refine the baseline model for each query cluster using a novel labeling and refinement strategy.

The key idea behind our approach is that the baseline model, while effective in capturing general semantic relationships, may not be sensitive to the specific characteristics of different query clusters. By refining the model for each cluster, we can better capture the nuances of various query types and provide more accurate and personalized retrieval results.

In summary, our cluster-based language model for e-commerce retrieval and ranking leverages the power of pre-trained language models and query clustering to deliver more accurate and personalized product retrieval results. By adapting the model to different query types, we can address the diverse needs of users in large-scale e-commerce environments and improve overall platform performance.

## 2 Related Work

The field of e-commerce retrieval and ranking has seen significant advancements over the years, with various techniques being proposed and developed. Our proposed cluster language model builds upon the successes of these existing techniques and introduces a novel approach to improve e-commerce retrieval performance. The most notable related work includes:

Learning to Rank: Learning to Rank (LTR) models are supervised machine learning techniques designed to optimize the ranking of items based on relevance. These approaches include pointwise, pairwise, and listwise ranking methods. Our proposed method differs from traditional LTR models by utilizing pre-trained language models and clustering queries to better capture the semantic relationships between queries and items (Burges, 2010; Freund et al., 2003).

Vector Space Models: Traditional information retrieval models, such as the Vector Space Model (VSM), utilize techniques like TF-IDF and cosine similarity to rank documents based on their relevance to a given query. Our approach enhances this concept by leveraging pre-trained language models and query clustering to better represent the semantic space and improve ranking performance (Salton et al., 1975).

Latent Semantic Analysis: Latent Semantic Analysis (LSA) captures the semantic relationships between queries and items for improved retrieval and ranking. Our method extends this idea by incorporating pre-trained language models and query clustering to further refine the semantic understanding of queries and items, leading to more accurate retrieval results (Deerwester et al., 1990).

Neural IR Models: Deep learning-based models, such as Convolutional Neural Networks (CNNs) and Recurrent Neural Networks (RNNs), have been applied to information retrieval tasks for text and image-based representations of items. Our proposed model takes advantage of the powerful representation capabilities of pre-trained language models and query clustering to improve the retrieval and ranking performance for e-commerce (Huang et al., 2013; Palangi et al., 2015).

Pre-trained Language Models: The use of pre-trained language models, such as BERT, GPT, and RoBERTa, has gained popularity in recent years for various natural language processing tasks, including e-commerce retrieval and ranking. Our approach differs from existing pre-trained language model applications by introducing query clustering and model refinement for each query cluster, which enhances the model's ability to capture the nuances of different query types and provide more personalized retrieval results (Devlin et al., 2018; Radford et al., 2018; Liu et al., 2019).

By combining the strengths of pre-trained language models, query clustering, and model refinement, our proposed cluster language model addresses the challenges of delivering accurate and personalized product retrieval results in large-scale e-commerce environments.

## 3 Methodology

In this section, we describe the technical details of our baseline model which is one of the top in-house models in terms of recall, NDCG (Wang et al., 2013) and execution time. Also, we present the proposed method that leverages the retrieval of the fine-tuned baseline model.

### 3.1 Sentence Transformer Architecture

The baseline model is essentially a sentence transformer, (Reimers & Gurevych, 2019) that is based on a bi-encoder architecture that contains two DistilBERT models (SanhSanh et al., 2019). These models are identical and share the same weights. The DistilBERT is a transformer-based model with 6 layers of self-attention and feed-forward neural networks. Each attention layer contains 12 attention heads. The self-attention mechanism allows the model to attend to different parts of the input sequence, allowing it to learn to represent different aspects of the input in different ways. In the context of sentence embeddings, the DistilBERT takes in a sentence as input and generates a (1, 768) size vector representation of that sentence. This vector representation is the sentence embedding used to measure the semantic similarity between two sentences or to classify a sentence into one of several categories.

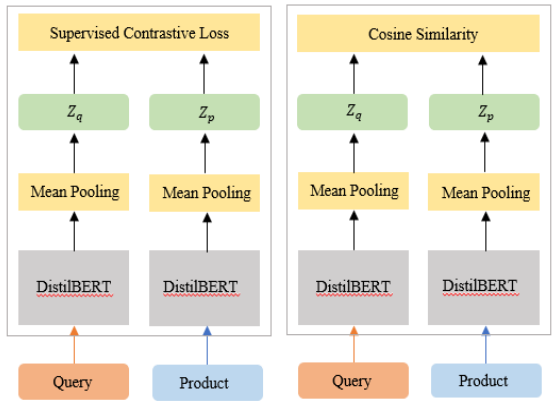

Figure 1: (Left) The architecture of the baseline model for training. (Right) The architecture of the baseline model for inference

Figure 1 presents our baseline model. It processes query (*q*) and product (*p*) sentences as pairs during training and testing. The baseline model adds a mean pooling operation to the output of the [CLS] token of the DistilBERT. In training, the embeddings of *q* and *p* denoted by $Z_q$ and $Z_p$ respectively, are used to optimize the supervised contrastive loss (Hadsell et al., 2006). In inference, the cosine similarity between $Z_q$ and $Z_p$ for different *p* is computed and based on that value, each *p* can be ranked with respect to *q*.

**Pretraining the Baseline Model:** We selected a pretrained sentence transformer model: MS MARCO-DistilBERT-Base-v2 from Hugging Face (*Hugging Face Transformers Library*, 2021). This model is built on a variant of the DistilBERT model, which was pre-trained using a large corpus of text data including the MS MARCO dataset (Nguyen et al., 2016; Wolf et al., 2020). The pre-training was done using masked language modeling (MLM) and next-sentence prediction (NSP) tasks. The MS MARCO-DistilBERT-Base-v2 model was further fine-tuned on the MS MARCO Passage Ranking task (*Hugging Face Transformers Library*, 2021), which is a large-scale information retrieval task that involves ranking a set of passages based on their relevance to a given query. The fine-tuning process involves training the model to predict the relevance score of given passage to a given query.

During the fine-tuning process, the model is trained using a combination of cross-entropy loss and Mean Reciprocal Rank (MRR) loss (*Hugging Face Transformers Library*, 2021). The cross-entropy loss is used to optimize the model's probability distribution over the candidate passages, while the MRR loss is used to optimize the rank of the relevant passage in the list of candidate passages.

**Fine Tuning the Baseline Model:** The baseline sentence transformer model is fine-tuned on query and product data in an e-commerce application. This task is performed by using a contrastive learning strategy (Hadsell et al., 2006). Let the training query set and the product set be *Q* and *P* respectively. For *q* ∈ *Q*, and *p, n* ∈ *P*, the input (*q, p*) is labeled with 1, and the input (*q, n*) is labeled with 0 considering that *p* is a positive sentence and *n* is a negative sentence (Hadsell et al., 2006). The goal of contrastive learning is to find parameters *W* of a family of functions *G*, to map a collection of high-dimensional inputs onto a low-dimensional manifold. For *x* = {*p, n*}, this mapping is such that the Euclidean distance between points on the manifold, given by: $D_w(Z_q, Z_x) = \|G_w(Z_q) - G_w(Z_x)\|^2$ closely approximates the semantic similarity of the inputs in the input space by minimizing the following objective function (Hadsell et al., 2006).

$$L = (1 - Y)\frac{1}{2}D_W^2 + Y\frac{1}{2}\{max(0, m - D_W)\}^2 \quad (1)$$

Where, *Y* = {0, 1}, and *m* > 0 is a margin. The fine-tuned baseline model can be used for inference. For our e-commerce use case, the retrieval set of interest is limited to the first 100 products since a typical customer is less likely to explore the search result beyond this limit. This set is known as the *Top Product Set* for the given query *q* and is denoted by $P_q$. Even though the baseline method can search for products with a competitive recall@24 on unseen queries, its performance at smaller retrieval sets is observed to be relatively weak.

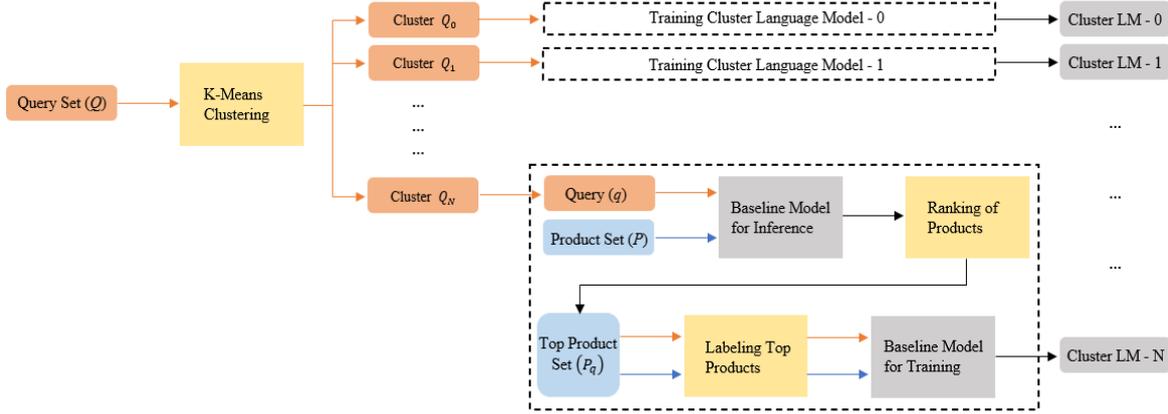

Figure 2: The architecture of the proposed Cluster Language Model for training

## 3.2 The Proposed Method

Although the bi-encoder architecture is considered fast, its accuracy is often compromised. This is an inherent drawback of the baseline model. The intention of the proposed method is to come up with a solution and enhance the product search up to @24 with a minimum of additional training. Thus, it provides an alternative approach to the popular bi-encoder and cross-encoder combination (Rosa et al., 2022; Ortiz-Barajas et al., 2022) in semantic search.

**Labeling Top Products for Each Query:** The rationale behind this step is the following observation: even though our baseline model is effective in capturing general semantic relationships, it may not be sensitive to the specific characteristics of different query clusters. To mitigate this problem, we introduce a novel labeling strategy for the elements of the Top Product Set and produce a new training dataset per query. This dataset will be used to train the corresponding cluster language model, depending on the query cluster where the training query is located.

For a given query $q$, the *Top Product Set* can be presented as $P_q = \{p_1, p_2, ..., p_{99}\}$. In an ideal retrieval, the order of the elements of the Top Product Set is such that all the purchased products should appe added-to-cart products.

However, in real retrievals, some of the purchased products are often seen to be forced back by unpurchased (impressed or added-to-cart) products due to the extreme similarities between them. To create the new training dataset for the corresponding cluster language model, we pair each query $q$ with some elements of its *Top Product Set* $P_q$ and label them using the following rule.

Let the query $q$ be an arbitrary query; power wash cleaner, and identify the last purchased product for $q$ in its $P_q$. Let this last purchased product be $p_k$. Now, for all $i < k$, we label query-product pairs $(q, p_i)$ with 0 if $p_i$ is not a purchased product. Also, we label all query-product pairs $(q, p_i)$ with 1 if $p_i$ is a purchased product, as shown in Table 1. The above relabeling helps the baseline model specifically suppress the unpurchased products that are more similar to the purchased products.

According to Table 1, a given product is identified by using its PRODID which is a unique identifier. The attribute 'Type' is used only to indicate whether the product is purchased (P) or unpurchased (U). The unlabeled query-product pairs are discarded.

**Generating Query Clusters:** The training of the proposed cluster language model comprises two phases as shown in Figure 2. In the first phase, we split the queries into $N$ clusters using K-Means clustering algorithm. For this task, the K-Means algorithm is trained on the embedding $Z_q$ for all $q \in Q$ by using the baseline model for inference (without computing cosine similarity).

| Rank | Query | Product | Type | Label |
|---|---|---|---|---|
| 0 | power wash cleaner | P42710 | P | 1 |
| 1 | power wash cleaner | P43322 | U | 0 |
| 2 | power wash cleaner | P51270 | U | 0 |
| … | … | … | … | … |
| k | power wash cleaner | P52993 | P | 1 |
| … | … | … | … | … |
| 99 | power wash cleaner | P58671 | U | na |

Table 1: The proposed training dataset of the cluster language model for a selected query.

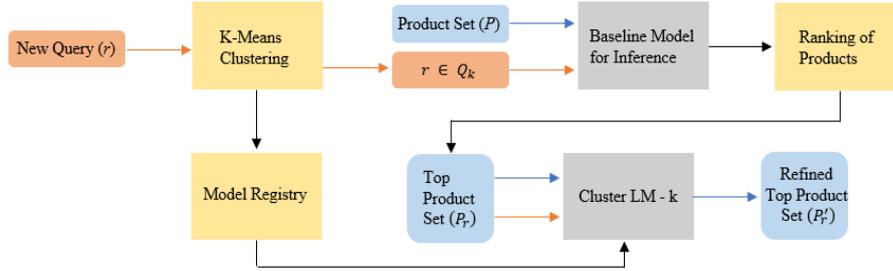

Figure 3: The inference pipeline of the proposed Cluster Language Model

This process generates clusters of queries that are more semantically similar. Let the $j$ th cluster of $Q$ be denoted by $Q_j$, where $j$ is known as the cluster ID. Then in each cluster, we store only the actual queries. The value of $N$ should not be either very small or very large and is determined by investigating the performance of clustering. In the ideal case, $N$ results in query clusters in which the within-cluster variance is much less compared to the between-cluster variance.

**Training Cluster Language Model on Individual Clusters:** The second phase of training the proposed cluster language model is implemented recursively on each cluster. This process is presented by the workflow illustrated for the query cluster $Q_N$ in Figure 2, where for each query $q$, we first rank the product embeddings with respect to the query embedding and obtain the Top Product Set $P_q$ for using the baseline model for inference. Then we label the top products as discussed to generate a new training dataset for each query as shown in Table 1. For all the queries in the given cluster, we fine tune the baseline model using the above mini datasets and the optimization presented in Chapter 3.1 to generate the Cluster LM – $N$. After the second phase is completed for all the clusters, the collection of Cluster LM – $k$, where $0 \leq k \leq N$ known as the Cluster Language Model is stored in a model registry.

Since the baseline model parameters have been generally tuned for all queries and products, the proposed cluster-based fine tuning helps the model parameters learn local manifolds in the feature space efficiently. Therefore, the clustered queries help better capture the nuances of various query types within that cluster. As a result, the relevance of the retrieved products will be improved.

**Inference Using Cluster Language Model:** In inference, a new query is assigned to the respective cluster by the trained K-Means algorithm. Thus, receiving the cluster ID for the new query $r$, we can select the corresponding Cluster LM from the model registry for inference. First, the new query $r$ and the product set $P$ is used to generate the *Top Product Set $P_r$* as shown in the inference pipeline in Figure 3. Then we input $P_r$ (completely) to the inference architecture of the selected Cluster LM to generate a *Refined Top Product Set $P_r'$*.

In general, the *Refined Top Product Set* is a better retrieval than the *Top Product Set* for new queries of each cluster, providing more accurate and personalized retrieval results.

## 4 Experiments

In this section, we introduce our datasets, data preprocessing, evaluation process, and present results to demonstrate the impact of the proposed method in production.

### 4.1 Datasets

We used the same product catalog for training and validation of both the baseline and the proposed methods. This dataset consists of approximately 1M unique products. Each product contains only the following attributes.

- OMSID: the unique product identification number
- Title: the product name
- Brand: the product brand
- ColorFinish: the overall color of the product
- Leaf: the category of the first level of the hierarchical taxonomy

An example of a product sentence is: *P52993 electric pressure washer sun joe green pressure*

*washer*. For the training and validation of the baseline and the proposed methods, we used the same query sets. The queries are made of refined customer search phrases that are free of typos. For the training and validation, we used about 15.4M unique queries and 24K unique queries respectively.

### 4.2 Experiment Settings and Preprocessing

All experiments and preprocessing are conducted on the Google Cloud Platform using NVIDIA A100 GPU using Python 3 and PyTorch 1.11. Major libraries used: SentenceTrandformer (Thakur et al., 2020) from Hugging Face and MiniBatchKMeans from scikit-learn (Sculley, 2010). The training dataset consists of roughly 60M query-product pairs. Each query is paired with a relevant purchased product, impressed, or added-to-cart product. The max token length is 40. The training batch size is 256. The number of epochs used for the baseline model and the Cluster Language Model: 15 and 5 respectively.

To train the proposed method, the training queries are clustered by using K-Means clustering. According to our initial experiments based on the training query set, we observed that the Cluster LM – $k$ for $0 \leq k \leq N$, consistently fails to outperform the baseline model on the respective validation queries if the size of Cluster $Q_k$ is about 1M or more. Although the query clusters of that scale are hard to describe based on their contents, this observation was notable despite different cluster members. Thus, we can safely assume that the size of any query cluster should be much less than 1M. For our experiments, we satisfy the above condition by setting $N = 29$ and resulting in 30 clusters with a mean cluster size of 513K with a standard deviation of 180K.

However, a large number of clusters such as $N = 100$ could lead to memory-related issues when deploying the proposed model in the cloud. Also, it could lower the overall performance of the cluster language model as many smaller clusters may contain quite similar queries. This reduces the probability that a random query being assigned to the correct Cluster LM – $k$ for $0 \leq k \leq N$ during the inference. Figure 4 shows the Calinski Harabasz (Caliński & Harabasz, 1974) Score Elbow for K-Means Clustering and according to which the elbow occurs when the number of clusters is 2 ($k = 3$). Thus, a larger number of clusters greater would produce weaker clusters of queries.

### 4.3 Evaluation Process

Our evaluation process is conducted to measure the following two tasks.

**Matching:** the intention of this task is to retrieve all the relevant purchased products for a given query. We use recall to measure matching at different retrieval thresholds such as 1, 2, 4, 8, 12, 24, and 100.

$$Recall@k = \frac{\#\ of\ purchased\ products\ @k}{\min(k, Total\ \#\ of\ purchased\ products)} \quad (2)$$

**Ranking:** the intention of this task is to order the retrieved products by their relevance. We use Normalized Discounted Cumulative Gain (NDCG) (Wang et al., 2013) to measure ranking at the retrieval thresholds mentioned above. We used 1 to denote a product if it is purchased and 0 to denote the product otherwise and applied the following definitions.

$$NDCG@k = \frac{DCG@k}{IDCG@k} \quad (3)$$

Where $DCG@k = \sum_{i=1}^{k} product_k / \log_2(k+1)$, and IDCG@k is the Ideal Discounted Cumulative Gain at k.

### 4.4 Results

Our experiment were conducted to evaluate the effectiveness of the baseline model and the proposed Cluster Language Model in retrieving relevant purchased products from our product catalog. The overall performance of both models in matching and ranking is presented in Table 2. These results show that the Cluster Language Model has a significantly higher recall rate than that of the baseline model up to the retrieval threshold@8. At higher thresholds, both models have quite similar matching performances. However, in ranking, the Cluster Language Model demonstrates a great improvement at every retrieval threshold.

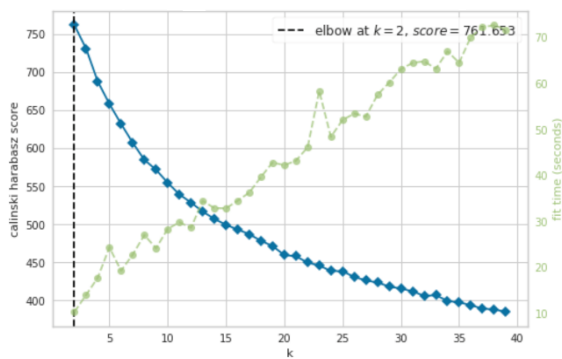

Figure 4: Calinski Harabasz Score Elbow for K-Means Clustering.

| Retrieval Threshold | Recall | | NDCG | |
|---|---|---|---|---|
| | Baseline Model | Cluster LM | Baseline Model | Cluster LM |
| @1 | 47.4 | 53.9 | 48.2 | 54.8 |
| @2 | 52.6 | 56.7 | 51.6 | 56.7 |
| @4 | 61.8 | 63.6 | 56.3 | 60.2 |
| @8 | 73.1 | 73.2 | 61.1 | 64.1 |
| @12 | 79.3 | 78.9 | 64.8 | 67.5 |
| @24 | 87.9 | 87.9 | 66.2 | 69.0 |
| @100 | 96.7 | 96.7 | 68.4 | 71.2 |

Table 2: The overall performance of baseline model and Cluster Language Model

The retrieval threshold@24 is considered to be a benchmark for our product search tasks. We investigated the cluster-level performance of the above two models and presented them in Figure 5. Also, we identified seven clusters in which the difference between the performance of the two models in Recall@24 is greater than 2%. Out of these cases, in four cases, the Cluster Language Model leads as shown in Table 3 and in the other three cases, the baseline model leads as shown in Table 4.

For each cluster (denoted by ID), Table 3 presents the percentage of training queries used in each cluster. It also provides the frequency (as a percentage) at which the purchased products of the testing dataset appear in the training dataset.

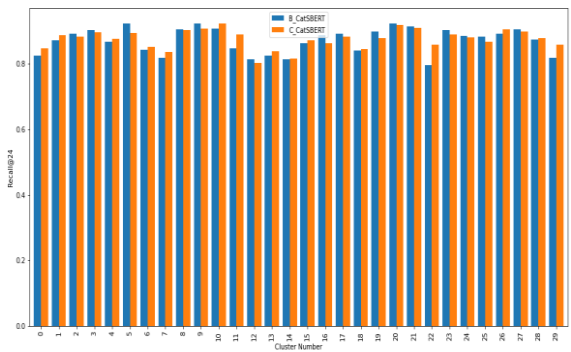

Figure 5: The cluster-level performance of the baseline model (blue) and the Cluster Language Model (orange) in matching, measured for Recall@24

Generally, we can expect this measure to be higher for the clusters in which the proposed method performs well during testing. The rationale behind this idea is that the language models tend to bias toward the majority of data (Wolfe & Caliskan, 2021).

The mean L2 distance shown in Table 3 measures how far a given cluster member is from the cluster center on average. Thus, the lower mean L2 distance implies a denser cluster. Table 4 presents a similar analysis for the clusters in which the baseline model outperforms the proposed method

| ID | Training Data (%) | Occurrence of Testing Purchased Products in Training Data (%) | Mean L2 Distance | Recall@24 | |
|---|---|---|---|---|---|
| | | | | Baseline Model | Cluster LM |
| 0 | 2.44 | 2.95 | 1.051 | 82.4 | 84.7 |
| 11 | 3.13 | 2.51 | 1.054 | 84.8 | 89.1 |
| 22 | 2.53 | 1.29 | 1.055 | 79.5 | 85.8 |
| 29 | 3.79 | 3.16 | 1.038 | 81.9 | 85.9 |

Table 3: The best matching for the Cluster Language Model at cluster-level

| ID | Training Data (%) | Occurrence of Testing Purchased Products in Training Data (%) | Mean L2 Distance | Recall@24 | |
|---|---|---|---|---|---|
| | | | | Baseline Model | Cluster LM |
| 5 | 3.48 | 3.29 | 1.049 | 92.3 | 89.6 |
| 9 | 3.22 | 2.15 | 1.054 | 92.4 | 90.3 |
| 16 | 1.51 | 0.78 | 1.083 | 88.7 | 86.2 |

Table 4: The best matching for the baseline model at cluster-level

Figure 6 illustrates the relative L2 distance between cluster centers as a heatmap. According to this plot, the centers of clusters 3, 16, and 17 are located relatively further away from the rest of the cluster centers. Conversely, the centers of clusters 10, 15, and 19 are much close to the rest of the cluster centers. Having more distinct cluster centers helps the K-Means algorithm assign new queries to the respective clusters rather correctly.

In addition, we measured and compared the execution time for both models. Table 5 shows the average time to process a single query by using these methods. The Cluster Language Model needs to complete all the processes shown in Table 5. Thus, it takes about 49.5 ms to completely process a given query. The baseline model only needs the second and third processes. Therefore, it is about ten times as fast as the proposed method.

| Process | Number of Queries Used | Total Time (s) | Time to Process a Single Query (ms) |
|---|---|---|---|
| Assign queries to Cluster LM using K-Means lustering | 24120 | 0.085 | 0.003 |
| Produce query embeddings using baseline model | 890 | 0.297 | 0.333 |
| Search for 100 matching products for each query | 890 | 3.417 | 3.839 |
| Produce embeddings for *Top Product Set*, search for matching products, and obtain *Refined Top Product Set*. | 890 | 40.348 | 45.335 |

Table 5: The execution time of the baseline model and the Cluster Language Model

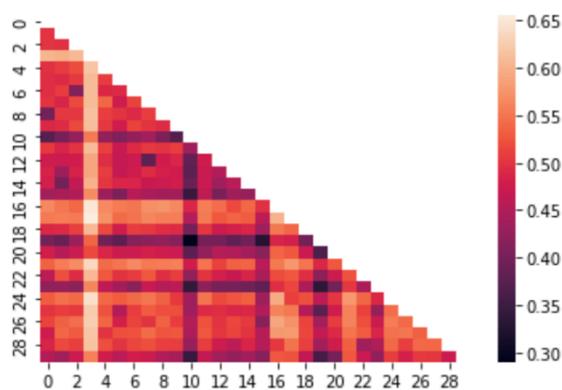

Figure 6: The relative L2 distance between cluster centers

## 5 Conclusion

In conclusion, this paper presented a novel approach to enhance the product search performance of the bi-encoder architecture by introducing a cluster-based fine-tuning method. The proposed method demonstrated significant improvement in recall rates up to the retrieval threshold@8, and consistently better-ranking performance across all thresholds, compared to the baseline model. Despite the increased processing time for the Cluster Language Model, it offers an alternative method to the popular bi-encoder based retrieval models in semantic search, addressing the inherent accuracy trade-offs often faced by the baseline model. The cluster-level analysis revealed that the proposed method performs well in denser clusters with a higher frequency of testing purchased products appearing in the training data. Although the baseline model outperforms the proposed method in certain clusters, the overall performance of the Cluster Language Model is superior. The L2 distance heatmap provides insights into the distinctiveness of cluster centers, which helps in the correct assignment of new queries to respective clusters.

This study demonstrates the potential of leveraging clustering techniques and fine-tuning to enhance semantic search in e-commerce applications. Further research could explore other clustering algorithms, the use of additional features, and optimization strategies to improve the performance of the proposed method and reduce processing time.

## Acknowledgments

This work was sponsored by The Home Depot. The opinions, findings, and conclusions or recommendations expressed in this material only reflect those of the authors in their individual capacities.